\documentclass[english]{article}
\usepackage{jheppub}
\renewcommand{\Re}{\operatorname{Re}}
\renewcommand{\Im}{\operatorname{Im}}
\newcommand{\sign}{\operatorname{sign}}
\newcommand{\arctanh}{\operatorname{arctanh}}
\allowdisplaybreaks

\title{One-loop pentagon integral in $d$ dimensions from differential equations in $\epsilon$-form.}

\author[a,b]{Mikhail~G.~Kozlov}
\author[a]{Roman~N.~Lee}

\affiliation[a]{Budker Institute of  Nuclear Physics, Novosibirsk, 630090 Russia}
\affiliation[b]{Novosibirsk State University, Novosibirsk, 630090 Russia}

\emailAdd{m.g.kozlov@inp.nsk.su}
\emailAdd{r.n.lee@inp.nsk.su}

\abstract{
	We apply the differential equation technique to the calculation of the one-loop massless diagram with five onshell legs. Using the reduction to $\epsilon$-form, we manage to obtain a simple one-fold integral representation exact in space-time dimensionality. The expansion of the obtained result in $\epsilon$ and the analytical continuation to physical regions are discussed.
}
\begin{document}
	\maketitle

\section{Introduction}

One-loop multi-leg diagrams are the building blocks for the construction of the next-to-leading order (NLO) amplitudes in the Standard Model and beyond. Within the standard approach, based on IBP reduction, these diagrams are expressed in terms of the one-loop master integrals. Scalar pentagon integral is somewhat special among them because it is the last and the most complicated piece needed for calculations of NLO multi-particle amplitudes with external legs lying in four-dimensional linear space. 

Another reason to study one-loop pentagon integral is the Bern--Dixon--Smirnov (BDS) ansatz~\cite{BDS:2005}. This ansatz relates MHV multiloop amplitudes in the planar limit of $\mathcal{N}=4$ supersymmetric Yang-Mills theory to the one-loop amplitude with the same number of legs. The ansatz is violated
for amplitudes with more than five legs, therefore, the five-leg amplitudes are the most complicated ones which satisfy the ansatz. The massless pentagon integral in $d=4-2\epsilon$ also appears in the calculation of the Regge vertices for the multi-Regge processes of QCD in the next-to-leading order~\cite{FF:1998}. The one-gluon production vertex in the NLO must be known at arbitrary $ d $ for the calculation  of the NLO Balitskii--Fadin--Kuraev--Lipatov (BFKL)~\cite{FF:2005} and Bartels--Kwiecinski--Praszalowicz (BKP)~\cite{BFLV:2012} kernels.

In the present paper we consider the one-loop pentagon integral with massless internal lines and onshell external legs, which we call below \textit{the pentagon integral} for brevity. In Ref. \cite{BDK:1994}, it was shown that through $\epsilon^0$ order the pentagon integral in $d=4-2\epsilon$ dimensions can be expressed via the box integrals with one offshell leg.  However, deriving higher orders in $\epsilon$ appeared to be a much more difficult task. In Ref. \cite{BDK:1994}, it was shown that higher-order terms are related to the expansion of the same pentagon integral in $6-2\epsilon$ dimensions. In Ref. \cite{DDGS:2010} the Regge limit of the pentagon integral in $6-2\epsilon$ dimensions was considered. The coefficients of expansion through $\epsilon^2$ were presented in terms of the Goncharov's polylogarithms. In Ref. \cite{Tarasov:2010} a rather complicated representation for the pentagon integral has been obtained using dimensional recurrence relation \cite{DerkachovHonkonenPismak1990,Tarasov1996}. The integral was expressed in terms of the Appell function $F_3$ and hypergeometric functions $_pF_q$. The expression was obtained for the region where all kinematic variables were negative and ordered in a specific way.

In a sense, the goal of the present paper is the same as that of Ref. \cite{Tarasov:2010}, but the method is different and the result obtained is strikingly simple, see Eq. \eqref{eq:pentagon_final}. We apply the approach first introduced in Ref. \cite{H:2013}, based on the reduction of the differential equations for master integrals to the Fuchsian form with factorized dependence of the right-hand side on $\epsilon$ ($\epsilon$-form). If this form is achieved simultaneously for the differential systems with respect to all variables, it is automatically possible to rewrite these systems in a unified $d\log$ form, which essentially simplifies the search for the solution. After finding $d\log$ form we choose not to follow conventional strategy of finding $\epsilon$-expansion order by order, but to obtain the result exactly in the dimension of space-time. The result appeared to have a remarkably simple form and provides a one-fold integral representation of arbitrary order of $\epsilon$ expansion `out-of-the-box'. Firstly we consider the integral in Euclidean region and then perform the analytical continuation to all other regions with real kinematic invariants.

\section{Definitions and result}

The pentagon integral is defined as 
\begin{equation}
P^{\left(d\right)}\left(s_{1},\,s_{2},\,s_{3},\,s_{4},\,s_{5}\right)=\int\frac{d^{d}l}{i\,\pi^{d/2}\prod_{n=0}^{4}\left(l_{n}^{2}+i0\right)}\,,
\end{equation}
where 
\begin{equation}
l_{n}=l-\sum_{i=1}^{n}p_{i}\,,
\end{equation} and $p_i$ are the incoming momenta,
\begin{equation}
	p_{i}^{2}=0\,,\quad\sum_{i=1}^{5}p_{i}=0\,,
\end{equation}
and the invariants $s_i$ are defined as
\begin{equation}
s_{n}=2p_{n-2}\cdot p_{n+2}\,.	
\end{equation}
Here and below we adopt cyclic convention for indices, e.g. $s_{n\pm5}=s_{n}$. 
We introduce the following notation
\begin{equation}
r_{n} =\sum_{i=0}^{4}(-1)^{i}s_{n+i}s_{n+i+1}\,,\quad
\Delta =\det\left(2p_{i}\cdot p_{j}|_{i,j=1,\ldots4}\right)=\sum_{i=1}^{5}r_{i}r_{i+2}\,,\quad
S =4s_{1}s_{2}s_{3}s_{4}s_{5}/\Delta\,.
\end{equation}

Using techniques described in detail in the succeeding sections, we obtain the following exact in $d$ representation for $P^{\left(6-2\epsilon\right)}$ for real $s_{i}$ (of arbitrary signs)
\begin{multline}\label{eq:pentagon_final}
P^{\left(6-2\epsilon\right)}\left(s_{1},\,s_{2},\,s_{3},\,s_{4},\,s_{5}\right)=\frac{C(\epsilon)}{\epsilon}\Bigg[
\Theta\left(s_is_j>0\right)\frac{2\pi^{3/2}\Gamma\left[1/2-\epsilon\right]}{\Gamma\left[1-\epsilon\right]\sqrt{\Delta}}\left(-S-i0\right)^{-\epsilon}
\\
+\sum_{i=1}^{5}\left(-s_{i}-i0\right)^{-\epsilon}\intop_{1}^{\infty}\frac{dt}{t}t^{\epsilon}
\Re\ \frac1{b_i(t)}\bigg\{
\arctan\frac{b_{i}(t)}{r_{i}}-\arctan\frac{b_{i}(t)}{r_{i+2}}-\arctan\frac{b_{i}(t)}{r_{i-2}}\\
+\frac{\pi}{2}\left[\sign r_{i+2}+\sign r_{i-2}-\sign r_{i} -\sign\left(r_{i+2}+r_{i-2}\right)\right]
\bigg\}\Bigg]\,,
\end{multline}
where $b_i(t)=\sqrt{\left(St/s_{i}-1\right)\Delta + i0}$ (obviously, $+i0$ can be replaced by $-i0$), 
\begin{equation}
C(\epsilon)=\frac{2\Gamma(1-\epsilon)^{2}\Gamma(1+\epsilon)}{\Gamma(1-2\epsilon)}\,,
\end{equation}
and $\Theta\left(s_is_j>0\right)$ equals to $1$ if \textbf{all} $s_i$ are of the same sign, and zero otherwise. By $\Re \frac{\arctan(\sqrt{x+i0}/r)}{\sqrt{x+i0}}$ we understand the function
\begin{equation}
f(x,r)=\begin{cases}
\frac{\arctan(\sqrt{x}/r)}{\sqrt{x}}\,,& x>0\\
\frac1{2\sqrt{-x}}\log\left|\frac{r+\sqrt{-x}}{r-\sqrt{-x}}\right|\,,& x<0
\end{cases}
\end{equation} 

The $\epsilon$-dependence in the integrand of \eqref{eq:Pres} is confined to the factor $t^{\epsilon}$. Therefore, any order of $\epsilon$-expansion can be trivially written as a one-fold integral of elementary functions. In the Appendix \ref{sec:expansion} we explain how to rewrite this integral in terms of the Goncharov's polylogarithms. We  also demonstrate the cancellation of $O(\epsilon^{-1})$ terms.

In order to crosscheck our result, we have performed comparison with the numerical results for pentagon obtained using \texttt{Fiesta} 3, Ref. \cite{Fiesta3}, and found perfect agreement. Some results of the comparison are presented in Table \ref{tab:Fiesta}.
\begin{table}
	\small
	\begin{tabular}{|c|c|c|c|c|c|c|}
		\hline
		$ s_1 $ & $ s_2 $ & $ s_3 $ & $ s_4 $ & $ s_5 $ &  Our result \eqref{eq:pentagon_final_expansion} & \texttt{Fiesta} 3\\
		\hline
		0.331 & 0.846 & 0.346 & 0.512 & 0.243  & 
		\begin{tabular}{l}
			$ -15.20480 $ \\
			$\mspace{20mu} +[-69.2882-i47.767]\epsilon $
		\end{tabular} &
		\begin{tabular}{l}
			$ -15.2046(3)+i0.000(2)$ \\
			$\mspace{20mu}+[-69.288(1)-i47.77(1)]\epsilon$
		\end{tabular}\\
		\hline
		0.899 & 0.068 & 0.455 & 0.253 & -0.478 &   
		\begin{tabular}{l}  
			$ -2.191+i33.760$ \\ $\mspace{20mu}+[-100.38+i146.37]\epsilon  $ 
		\end{tabular}& 
		\begin{tabular}{l}
			$ -2.12(9)+ i33.77(9) $ \\ $\mspace{20mu}+[-100.0(6)+i146.8(6)]\epsilon$
		\end{tabular}\\
		\hline
		0.294 & 0.716 &  0.467 & -0.109 &  -0.552 &
		\begin{tabular}{l}  
			$ 17.759+ i34.223$\\
			$\mspace{20mu}+[ 21.331+i165.766 ]\epsilon$
		\end{tabular}& 
		\begin{tabular}{l}
			$ 17.9(1)+i34.3(1)$\\
			$\mspace{20mu}+[22.4(6)+i166.2(6)]\epsilon$
		\end{tabular}\\
		\hline
		0.317 & 0.932 & -0.233 & 0.206 & -0.114 & 
		\begin{tabular}{l}  
			$ 29.001+i35.675$ \\ $\mspace{20mu}+[77.733+i208.091]\epsilon$
		\end{tabular} & 
		\begin{tabular}{l}
			$ 29.2(1)+i35.7(1)$ \\ $\mspace{20mu}+[79(1) +i 209(1) ]\epsilon $
		\end{tabular}\\
		\hline
		0.036 & 0.573 & -0.896 & -0.467 &  -0.753 &
		\begin{tabular}{l}  
			$ 9.005- i25.175$ \\ $\mspace{20mu}+[85.619-i92.931]\epsilon $
		\end{tabular} & 
		\begin{tabular}{l}
			$ 9.05(6)-i25.16(6)$ \\ $\mspace{20mu}+[86.1(4)-i93.0(4)]\epsilon $
		\end{tabular}\\
		\hline
		-0.007 & -0.254 & 0.241 & -0.056 &  0.545 &
		\begin{tabular}{l}  
			$ 206.941-i44.552$ \\
			$\mspace{20mu}+[1246.582 -i119.587]\epsilon$
		\end{tabular} & 
		\begin{tabular}{l}
			$ 208(1) -i45(1) $\\
			$\mspace{20mu}+[1253(8) - i118(8) ]\epsilon $
		\end{tabular}\\
		\hline
		0.629 & -0.973 & -0.155 & -0.219 & -0.452 &
		\begin{tabular}{l}
			$ -0.0835-i22.7366 $ \\  $ \mspace{20mu}+[8.7037-i96.9259]\epsilon $
		\end{tabular}&
		\begin{tabular}{l}
			$ -0.06(4)- i22.74(4)$ \\ $ \mspace{20mu}+[8.9(3)-i96.9(2)]\epsilon  $
		\end{tabular}\\
		\hline
		-0.164 & -0.792 & -0.312 & -0.753 & -0.590 & 
		\begin{tabular}{l}  
			$ -12.731$ \\ $ \mspace{20mu}-57.351\epsilon $
		\end{tabular} &
		\begin{tabular}{l}  
			$ -12.731(3)-i0.000(2)$ \\ $\mspace{20mu}+[-57.35(1) - i0.00(1)]\epsilon  $
		\end{tabular}\\
		\hline
	\end{tabular}
	\caption{Comparison of the $\epsilon$-expansion of  $e^{\gamma\epsilon}P^{(6-2\epsilon)}$ with numerical results obtained using \texttt{Fiesta} 3 ($\gamma=0.577\ldots$ is the Euler constant). Our result is obtained by numerical integration of Eq. \eqref{eq:pentagon_final_expansion}.}	\label{tab:Fiesta}	
\end{table}

\section{Differential equations in $\epsilon$-form}

In this section, unless the opposite is explicitly stated, we consider integrals in $d=4-2\epsilon$ dimensions in ``Euclidean'' region
\begin{equation}\label{eq:euclidean}
	s_1<0\,,\quad s_2<0\,,\quad s_3<0\,,\quad s_4<0\,,\quad s_5<0\,.
\end{equation}

\begin{figure}
\centering
\includegraphics[width=\linewidth]{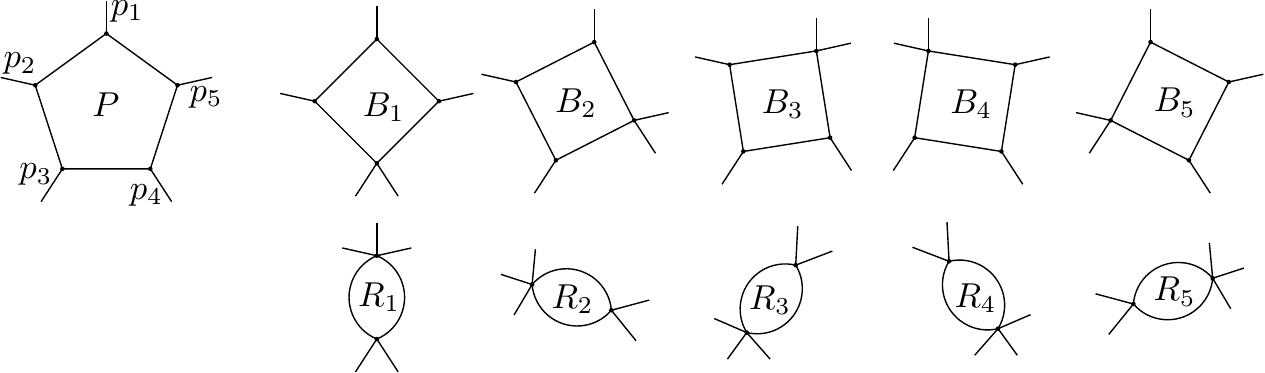}
\caption{Pentagon, box and bubble integrals.}
\label{fig:diagrams}
\end{figure}

We use IBP reduction, as implemented in  \texttt{LiteRed} package, Ref. \cite{LiteRed}, to obtain the system of partial differential equations for the pentagon integral $P$ and ten simpler master integrals, see Fig. \ref{fig:diagrams}. 

Introducing the column-vector 
\begin{equation}\label{eq:oldMIs}
{\boldsymbol J}=(P,\,B_1,\,B_2,\,B_3,\,B_4,\,B_5,\, R_1,\,R_2,\,R_3,\,R_4,\,R_5)^T\,,
\end{equation}
we may represent the system in the matrix form
\begin{equation}
\frac{\partial}{\partial s_i}{\boldsymbol J}=M_i(\boldsymbol{s},\epsilon){\boldsymbol J}\,,\quad i=1,\dots,5\,,
\end{equation}
where $M_i(\boldsymbol{s},\epsilon)$ are upper-triangular matrices of rational functions of $s_j$ and $\epsilon$. We benefit from knowing simpler masters, which are the bubbles
\begin{equation}
R_i=R(s_i)=\int\frac{d^{d}l}{i\,\pi^{d/2}l_{i+1}^2l_{i+3}^2}=\frac{C(\epsilon)}{2\epsilon(1-2\epsilon)}(-s_i)^{-\epsilon}
\end{equation} 
and the massless box integrals with one off-shell leg
\begin{equation}
B_i=B(s_{i+2},s_{i-2},s_i)
=\int\frac{d^{d}l}{i\,\pi^{d/2}\prod_{k=3}^{6}l_{i+k}^2}\,.
\end{equation}
The representation of the box integral in terms of the hypergeometric function obtained in Ref. \cite{BDK:1994} has the form
\begin{multline}\label{eq:box}
B(s_{i+2},s_{i-2},s_i)=\frac{C(\epsilon)}{\epsilon^2 s_{i+2}s_{i-2}}\biggl[(-s_i)^{-\epsilon}\,_2F_1\biggl(1,-\epsilon;1-\epsilon;1-\frac{(s_i-s_{i+2})(s_i-s_{i-2})}{s_{i+2}s_{i-2}}\biggr)\\- (-s_{i+2})^{-\epsilon}\,_2F_1\biggl(1,-\epsilon;1-\epsilon;1-\frac{s_i-s_{i+2}}{s_{i-2}}\biggr)-(-s_{i-2})^{-\epsilon}\,_2F_1\biggl(1,-\epsilon;1-\epsilon;1-\frac{s_i-s_{i-2}}{s_{i+2}}\biggr)\biggr]\,.
\end{multline}
This representation should be treated with care since the arguments of the hypergeometric functions may exceed $1$ and one must take care of direction the arguments approach the cut. One may check that the correct analytical continuation to the whole region $s_{i+2}<0,\,s_{i-2}<0,\,s_i<0$ is given by replacing in Eq. \eqref{eq:box} each $_2F_1(\alpha,\beta;\gamma;x)$ with $\Re {}_2F_1(\alpha,\beta;\gamma;x)=\frac12\sum_{\pm}{}_2F_1(\alpha,\beta;\gamma;x\pm i0)$. 

Next, we find appropriate basis in order to reduce the system to $\epsilon$-form, \cite{H:2013}. For our one-loop case the problem of finding the basis appears to be very simple and straightforward. In particular, we do not use much of the recipes given in Refs. \cite{eform:2015,HennLectures2014}. We do use though the basic idea of first reducing the diagonal blocks ($1\times1$) and then reducing the off-diagonal matrix elements. 
We end up with the basis $\widetilde{\boldsymbol J}=(\widetilde{P},\,\widetilde B_1,\ldots\widetilde R_5)^T\,,
$ which is related to \eqref{eq:oldMIs} as follows
\begin{gather}\label{eq:new-basis}
P=C(\epsilon)\frac{\sqrt{\Delta}}{s_1s_2s_3s_4s_5}\biggl(\widetilde{P}-\sum_{i=1}^5\frac{1}{2}\Bigl(1-\frac{r_i}{\sqrt{\Delta}}\Bigr)\widetilde{B}_i\biggr)\,,\\
B_i=\frac{C(\epsilon)}{s_{i+2}s_{i-2}}\widetilde{B}_i\,,\quad
R_i=\frac{C(\epsilon)\epsilon}{2(1-2\epsilon)}\widetilde{R}_i\,.
\end{gather}
Note that $\Delta>0$ in Euclidean region, so that $\sqrt{\Delta}$ is real.
The differential equations in the new basis can be written in $d\log$ $\epsilon$-form
\begin{align}
d\widetilde{P}=&-\epsilon\Biggl\{\widetilde{P} d\bigl(\log S\bigr)+\sum_{i=1}^5\Biggl[-\widetilde{B}_id\biggl(\log\Bigl(1+\frac{r_i}{\sqrt{\Delta}}\Bigr)\biggr)+\widetilde{R}_id\biggl(\log\frac{(\sqrt{\Delta}+r_i)(r_{i+2}+r_{i-2})}{(\sqrt{\Delta}+r_{i+2})(\sqrt{\Delta}+r_{i-2})}\biggr)\Biggr]\Biggr\}\,,\nonumber\\
d\widetilde{B}_i=&-\epsilon\biggl\{\widetilde{B}_id\biggl(\log\frac{s_{i-2}s_{i+2}}{s_{i-2}+s_{i+2}-s_{i}}\biggr)-\widetilde{R}_id\biggl(\log\frac{(s_i-s_{i-2})(s_i-s_{i+2})}{(s_{i-2}+s_{i+2}-s_i)s_i}\biggr)\nonumber\\
&+\widetilde{R}_{i-2}d\biggl(\log\frac{s_{i-2}-s_i}{s_{i-2}+s_{i+2}-s_i}\biggr)+\widetilde{R}_{i+2}d\biggl(\log\frac{s_{i+2}-s_i}{s_{i-2}+s_{i+2}-s_i}\biggr)\biggr\}\,,\nonumber \\
d\widetilde{R}_i =&-\epsilon\widetilde{R}_id(\log s_i)\,.\label{eq:system}
\end{align}
Let us now split the above differential system into five separate systems of dimension five. In view of possible further applications, we describe the splitting of sparse systems in some detail. Given a system $d\widetilde{\boldsymbol J}=dM\widetilde{\boldsymbol J}$ we schematically depict the matrix $dM$ by replacing each its nonzero element with ``$*$''. For the system \eqref{eq:system} we have
\begin{equation}
dM=\left[\tiny\begin{array}{*{11}c}
* &  * &  * &  * &  * &  * &  * &  * &  * &  * & *\\
0 &  * &  0 &  0 &  0 &  0 &  * &  0 &  * &  * & 0\\ 
0 &  0 &  * &  0 &  0 &  0 &  0 &  * &  0 &  * & *\\ 
0 &  0 &  0 &  * &  0 &  0 &  * &  0 &  * &  0 & *\\ 
0 &  0 &  0 &  0 &  * &  0 &  * &  * &  0 &  * & 0\\ 
0 &  0 &  0 &  0 &  0 &  * &  0 &  * &  * &  0 & *\\ 
0 &  0 &  0 &  0 &  0 &  0 &  * &  0 &  0 &  0 & 0\\ 
0 &  0 &  0 &  0 &  0 &  0 &  0 &  * &  0 &  0 & 0\\ 
0 &  0 &  0 &  0 &  0 &  0 &  0 &  0 &  * &  0 & 0\\ 
0 &  0 &  0 &  0 &  0 &  0 &  0 &  0 &  0 &  * & 0\\ 
0 &  0 &  0 &  0 &  0 &  0 &  0 &  0 &  0 &  0 & *
\end{array}
\right] \,.
\end{equation}
Then we interpret this schematic form as adjacency matrix of the directed graph, with ``$*_{ij}$'' denoting directed edge $ i \to j$. In general, the node $i$ is said to be an \textit{ancestor} of the node $j$ if there is a directed path from $i$ to $j$. A \textit{leaf} is a node which is not an ancestor of any other node. To each leaf we associate the subgraph consisting of the leaf itself and of all its ancestors.
For each such subgraph, we search for a solution of the original system having the form of the column vector with zeros put in all entries except the ones corresponding to the nodes of the subgraph. Then the general solution of the differential system is written as the sum over different leaves\footnote{The notion of a \textit{leaf} should be generalized in an obvious way in the case when some lowest non-zero sectors have several master integrals.}.

For our present case we have five leaves, $R_i;\  i=1,\ldots,5$. The subgraph of ancestors of $R_i$ contains $\widetilde{P},\,\widetilde B_i,\,\widetilde B_{i+2},\,\widetilde B_{i-2},\,\widetilde R_i$. In particular, for $i=1$ it means that we search for the solution in the form 
\begin{equation}
\widetilde{\boldsymbol J}^{(1)}=(\widetilde{P}^{(1)},\widetilde B_1^{(1)},0,\widetilde B_3^{(1)},\widetilde B_4^{(1)},0,\widetilde R_1^{(1)},0,0,0,0)^T\,.
\end{equation} 
Then the general solution is $\widetilde{\boldsymbol J}=\widetilde{\boldsymbol J}^{(1)}+\ldots+\widetilde{\boldsymbol J}^{(5)}$. Explicitly,
\begin{align}
\widetilde{P}&=\widetilde{P}^{(1)}+\widetilde{P}^{(2)}+\widetilde{P}^{(3)}+\widetilde{P}^{(4)}+\widetilde{P}^{(5)}\,,\\
\widetilde{B}_i&=\widetilde{B}_i^{(i)}+\widetilde{B}_i^{(i+2)}+\widetilde{B}_i^{(i-2)}\,,\\
\widetilde{R}_i&=\widetilde{R}_i^{(i)}\,.
\end{align}
From Eq. \eqref{eq:box} it is easy to identify functions $\widetilde{B}_i^{(i)}$, $\widetilde{B}_i^{(i+2)}$, and $\widetilde{B}_i^{(i-2)}$ as
\begin{align}\label{eq:simp-box}
\widetilde{B}_{i}^{(k)} &=\epsilon^{-2}(-1)^{(k-i)/2}(-s_k)^{-\epsilon}\Re{}_2F_1\biggl(1,-\epsilon;1-\epsilon;\frac{s_k}{S}\Bigl(1-\frac{r_i^2}{\Delta}\Bigr)\biggr)\,,\quad k=i\,,\;i\pm 2\,.
\end{align}
One can check explicitly that $\widetilde{B}_{i}^{(k)}$ satisfy required equations provided $d\log(s_i-s_{i\pm 2})$ is understood as  $(ds_i-ds_{i\pm 2})\mathcal{P}\frac{1}{s_i-s_{i\pm 2}}$. Here $\mathcal{P}\frac1x=\frac12\sum_{\pm}\frac1{x\pm i0}$ denotes the principal value prescription.

Using expression for $\widetilde{B}_{i}^{(k)}$ from \eqref{eq:simp-box} and the integral representation for hypergeometric function
\begin{equation}
\Re\,_2F_1(1,-\epsilon;1-\epsilon;x+ i0)-1=-\epsilon x\int_1^{\infty}\, dt\,t^{\epsilon-1}{\cal P}\frac{1}{t-x}\,,
\end{equation}
we arrive at the following differential equation for $\widetilde{P}^{(i)}$:
\begin{align}
&d\bigl((-S)^{\epsilon}\widetilde{P}^{(i)}\bigr)=H^{(i)}_ida_i+H^{(i)}_{i+2}da_{i+2}+H^{(i)}_{i-2}da_{i-2}\,,\label{eq:dSP}\\
&H^{(i)}_i=H^{(i)}_i(a_i,a_{i+2},a_{i-2})=-\Bigl(\frac{S}{s_i}\Bigr)^{\epsilon}\biggl[(1-a_i)\int_1^{\infty} {\cal P}\frac{dt\,t^{\epsilon-1}}{\frac{S}{s_i}t-1+a^2_{i}}\biggr]\,,\\
&H^{(i)}_{i\pm2}=H^{(i)}_{i\pm2}(a_i,a_{i+2},a_{i-2})=-\Bigl(\frac{S}{s_i}\Bigr)^{\epsilon}\biggl[(1-a_{i\pm2})\int_1^{\infty}{\cal P}\frac{dt\,t^{\epsilon-1}}{\frac{S}{s_i}t-1+a^2_{i\pm2}}-\frac{1}{\epsilon(a_{i+2}+a_{i-2})}\biggr]\,.
\end{align}
The right-hand side of Eq. \eqref{eq:dSP} depends only on three dimensionless variables $a_n=\frac{r_n}{\sqrt{\Delta}}$, ($n=i,\,i\pm2$). In particular, ${S}/{s_i}=1+a_{i-2}a_{i+2}-a_i(a_{i-2}+a_{i+2})$. It is easy to check that the right-hand side of \eqref{eq:dSP} is a total differential, i.e.,
\begin{equation}\label{eq:integrability}
\frac{\partial H^{(i)}_j}{\partial a_k}=\frac{\partial H^{(i)}_k}{\partial a_j}\,\quad (j,\,k=i,\,i\pm2)\,.
\end{equation}
Then from the differential equation $\frac{\partial}{\partial a_i}\bigl((-S)^{\epsilon}\widetilde{P}^{(i)}\bigr)=H^{(i)}_i$ we have
\begin{equation}\label{eq:Hsol}
(-S)^{\epsilon}\widetilde{P}^{(i)}=\int_{-\infty}^{a_i}H^{(i)}_i(a,a_{i+2},a_{i-2})da+g(a_{i+2},a_{i-2},\epsilon)\,,
\end{equation} 
where $g(a_{i+2},a_{i-2},\epsilon)$ is some function to be fixed. Using the equations $\frac{\partial}{\partial a_{i\pm 2}}\bigl((-S)^{\epsilon}\widetilde{P}^{(i)}\bigr)=H^{(i)}_{i\pm 2}$ and relation \eqref{eq:integrability}, it is easy to check that $g$ depends only on $\epsilon$. Indeed,
\begin{equation}
\begin{split}
&\frac{\partial}{\partial a_{i\pm 2}}\bigl((-S)^{\epsilon}\widetilde{P}^{(i)}\bigr)=\int_{-\infty}^{a_i}\frac{\partial H^{(i)}_i}{\partial a_{i\pm 2}}da_i+\frac{\partial g(a_{i+2},a_{i-2},\epsilon)}{\partial a_{i\pm2}}\,,\\
&\int_{-\infty}^{a_i}\frac{\partial H^{(i)}_i}{\partial a_{i\pm 2}}da_i=\int_{-\infty}^{a_i}\frac{\partial H^{(i)}_{i\pm 2}}{\partial a_{i}}da_i=H^{(i)}_{i\pm 2}-H^{(i)}_{i\pm 2}(a_i\rightarrow-\infty)=H^{(i)}_{i\pm 2}\,,
\end{split}
\end{equation} 
where we used the asymptotics $H^{(i)}_{i\pm 2}(a_i\rightarrow-\infty)=\epsilon^{-1} (-a_i)^{\epsilon}(a_{i-2}+a_{i+2})^{\epsilon-1}\to 0$. Therefore $\frac{\partial}{\partial a_{i\pm2}}g(a_{i+2},a_{i-2},\epsilon)=0$, or $g=g(\epsilon)$. Substituting the explicit form of $H^{(i)}_i$, we have
\begin{equation}
(-S)^{\epsilon}\widetilde{P}^{(i)}=-\int_{-\infty}^{a_i}(1-a)da\int_1^{\infty}\frac{dt}{t}\,\bigl(K(a)t\bigr)^{\epsilon}{\cal P}\frac{1}{K(a)t-1+a^2}
\end{equation}
where $ K(a)=1+a_{i-2}a_{i+2}-a(a_{i-2}+a_{i+2}) $. Note that $K(a)>0$ in the whole integration domain. 
Making the substitution $t \to t/K(a)$ and changing the order of integration we have
\begin{align}\label{eq:split2}
\widetilde{P}^{(i)}&=\widetilde{P}^{(i)}_0-\widetilde{P}^{(i)}_1 +(-S)^{-\epsilon}g(\epsilon)\\
\widetilde{P}^{(i)}_n&=-(-S)^{-\epsilon}\int_{\frac{S}{s_i}}^{\infty} t^{\epsilon-1}dt\int_{\frac{1-t+a_{i-2}a_{i+2}}{a_{i-2}+a_{i+2}}}^{a_i}da\ a^n{\cal P}\frac{1}{t-1+a^2}\notag\\
&=-(-s_i)^{-\epsilon}\sqrt{\Delta}\Re\int_1^{\infty}t^{\epsilon-1}dt\int_{r_i+2s_{i-1}s_{i+1}(1-t)}^{r_i}\frac{\bigl[r \Delta^{-\frac{1}{2}}\bigr]^ndr}{\Bigl(\frac{S}{s_i}t-1\Bigr)\Delta+r^2 + i0}\label{eq:Pki}\,.
\end{align}
It is remarkable that the integrals over $a$ and $t$ in $\widetilde{P}^{(i)}_1$ can be taken in terms of ${}_2F_1$. Moreover, it appears that $\widetilde{P}^{(i)}_1$ reduces to the sum of box functions $\tilde{B}_k^{(i)}$, Eq. \eqref{eq:simp-box}:
\begin{equation}\label{eq:P1B}
\widetilde{P}^{(i)}_1=\frac{1}{2}\Bigl(\widetilde{B}_{i}^{(i)}(\boldsymbol{s})+\widetilde{B}_{i+2}^{(i)}(\boldsymbol{s})+\widetilde{B}_{i-2}^{(i)}(\boldsymbol{s})\Bigr)\,.
\end{equation}
Hence, using equations~\eqref{eq:new-basis},~\eqref{eq:split2} and~\eqref{eq:P1B}, we can write the solution for pentagon integral in the form
\begin{equation}\label{eq:Pres}
P=C(\epsilon)\frac{\sqrt{\Delta}}{s_1s_2s_3s_4s_5}\Bigl(\sum_{i=1}^5\widetilde{P}^{(i)}_0+g(\epsilon)\,(-S)^{-\epsilon}\Bigr)+\sum_{i=1}^5\frac{r_i}{2s_{i-1}s_{i}s_{i+1}}B_i(\boldsymbol{s})\,,
\end{equation} 
In order to fix the constant $g(\epsilon)$, we notice that the condition $\Delta = 0$ implies the existence of linear relation between $p_1,\ldots, p_4$. Therefore, using partial fractioning, we can express the pentagon integral at $\Delta=0$ in terms of the box integrals. 
Moreover, $\Delta=0$ is not a branching point of $P$. The only way to satisfy these two conditions is to require that
\begin{equation}\label{eq:constraint}
\sum_{i=1}^5\widetilde{P}^{(i)}_0+g(\epsilon)\,(-S)^{-\epsilon}
\stackrel{\Delta\to 0}{\longrightarrow}0\,.
\end{equation} 
In order to calculate the limit $\Delta\to 0$ from within Euclidean region, we assume that $s_{2-5}$ are subject to the constraint $s_2s_3-s_3s_4+s_4s_5=0$. Then
\begin{equation}
 \Delta=s_1^2(s_2-s_5)^2+4s_1s_2s_5(s_3+s_4)\,.
\end{equation} 
In the limit $s_1\to 0$ we have
\begin{gather*}
\widetilde{P}_0^{(1)}\sim\widetilde{P}_0^{(2)}\sim\widetilde{P}_0^{(5)}\sim\Delta^{\frac{1}{2}-\epsilon}\to0\,,\\
\widetilde{P}_0^{(3)}\approx\widetilde{P}_0^{(4)}\to-\pi^{\frac32}\frac{\Gamma(1/2-\epsilon)}{\Gamma(1-\epsilon)}(-S)^{-\epsilon}\,.
\end{gather*} 
Therefore from Eq. \eqref{eq:constraint} we obtain
\begin{equation}\label{eq:g}
g(\epsilon)=2\pi^{\frac{3}{2}}\frac{\Gamma(1/2-\epsilon)}{\Gamma(1-\epsilon)}\,.
\end{equation}
Equations \eqref{eq:Pres} and \eqref{eq:g} determine $P^{(4-2\epsilon)}$. 

Let us consider now the dimensional recurrence relation 
\begin{equation}\label{eq:dre}
P^{(4-2\epsilon)}=\frac{\epsilon\Delta}{s_1s_2s_3s_4s_5} P^{(6-2\epsilon)}+\sum_{i=1}^5\frac{r_i}{2s_{i-1}s_is_{i+1}}B_i^{(4-2\epsilon)}\,,
\end{equation}
This relation is known since Ref.~\cite{BDK:1994} and can be routinely obtained with the \texttt{LiteRed}. Comparing \eqref{eq:dre} and \eqref{eq:Pres}, we obtain
\begin{equation}\label{eq:pentagon_euclidean}
P^{(6-2\epsilon)}=\frac{C(\epsilon)}{\epsilon\sqrt{\Delta}}\biggl[\sum_{i=1}^5\widetilde{P}^{(i)}_0+2\pi^{\frac{3}{2}}\frac{\Gamma(1/2-\epsilon)}{\Gamma(1-\epsilon)}(-S)^{-\epsilon}\biggr]\,.
\end{equation}

\section{Analytical continuation}
Let us now discuss the analytical continuation of the result obtained in the Euclidean region. The analytical continuation of a two-fold integral as a function of parameters is a highly nontrivial problem. Fortunately, the inner integral over $r$ in Eq. \eqref{eq:Pki} can be taken, and we represent $\widetilde{P}^{(i)}_0$ in the form
\begin{equation}\label{eq:PAi-F}
\Delta^{-\frac{1}{2}} \widetilde{P}^{(i)}_0=(-s_i)^{-\epsilon}\int_1^{\infty}dt\,t^{\epsilon-1}G_i(\boldsymbol{s},t)\,.
\end{equation}
The left-hand side of Eq. \eqref{eq:PAi-F}, including the factor $\Delta^{-1/2}$, is just the combination which enters Eq. \eqref{eq:pentagon_euclidean} and which requires the analytical continuation, and
\begin{multline}
\label{eq:G}
G_i(\boldsymbol{s},t)=\intop_{r_{i}+2s_{i+1}s_{i-1}\left(1-t\right)}^{r_{i}}dr\,\Re\left\{-\frac{1}{\left(St/s_{i}-1\right)\Delta+r^{2}+i0}\right\}\\
=-\frac{1}{2}\sum_{\pm}\Biggl\{\frac{1}{r_i}f\biggl(\frac{r_i^2}{\Delta\frac{S}{s_i}t-\Delta\pm i0}\biggr)-\frac{1}{r_i+2s_{i-1}s_{i+1}(1-t)}f\biggl(\frac{\bigl(r_i+2s_{i-1}s_{i+1}(1-t)\bigr)^2}{\Delta\frac{S}{s_i}t-\Delta\pm i0}\biggr)\Biggl\}\,.
\end{multline}
Here $f(z)=\sqrt{z}\arctan\left(\sqrt{z}\right)$ is a function defined on the complex plane with a cut going from $-\infty$ to $-1$. The Riemann surface, corresponding to the multivalued function $F(z)$ with the main branch defined by $F^{(0)}(z)=f(z)$, is glued of a set of sheets numbered by $n\in\mathbb{Z}$ with two cuts, one going from  $-\infty$ to $-1$ and the other going from $0$ to $\infty$. On the $n$-th sheet the function is defined as
\begin{equation}\label{eq:branches}
F^{(n)}(z)=-\frac{\sqrt{-z}}{2}\ln\frac{1+\sqrt{-z}}{1-\sqrt{-z}}+i\pi n\sqrt{-z}\,,\quad n\in\mathbb{Z}\,,
\end{equation}
where $\sqrt{\bullet}$ and $\ln(\bullet)$ denote the main branches of the corresponding functions. The gluing rules are
\begin{equation}\label{eq:Fn}
F^{(n)}(x\pm i0)=
\begin{cases}
\sqrt{x}\arctan(\sqrt{x})\pm\pi n\sqrt{x}=F^{(-n)}(x\mp i0)\,, & x>0\\
-\frac{1}{2}\sqrt{-x}\ln\frac{1+\sqrt{-x}}{\sqrt{-x}-1}+i\pi(n\pm 1/2)\sqrt{-x}=F^{(n\pm1)}(x\mp i0)\,,& x<-1\,.
\end{cases}
\end{equation}
The integrand of \eqref{eq:PAi-F} has the following branching points on the real axis of $t$:
\begin{itemize}
	\item $t=0$ is a branching point of the $ t^{\epsilon} $,
	\item $t_{ai}=1-\frac{(s_{i+2}-s_i)(s_{i-2}-s_i)}{s_{i+2}s_{i-2}}$, where the argument of the first function becomes $-1$,
	\item $t_{bi}=1-\frac{s_{i+2}-s_i}{s_{i+1}}$ and $t_{ci}=1-\frac{s_{i-2}-s_i}{s_{i-1}}$, where the argument of the second function becomes $-1$,
	\item $t_{0i}=1+\frac{r_i}{2s_{i+1}s_{i-1}}$, where the argument of the second function becomes $0$,
	\item $t_{\infty i}=\frac{s_i}{S}$, where arguments of both functions become $\infty$.
\end{itemize}
The sum over $\pm$  signs in Eq. \eqref{eq:G} translates into the sum over two different integration contours over $t$ in Eq. \eqref{eq:PAi-F}.

In general, the analytical continuation depends, in a highly non-trivial way, on the path in $\mathbb{C}^5$ space of $(s_1,\ldots,s_5)$ connecting a point in Euclidean region with the point of interest. However, the problem is essentially simplified if we restrict ourselves by the paths lying in the region $\mathcal{D}=\{\boldsymbol{s}|\Im s_i\geqslant0\}$. Using Feynman parametrization, it is easy to see that any two paths connecting a given pair of points and lying in $\mathcal{D}$ are equivalent. Therefore, the choice of a convenient path is totally in our hands provided that it lies in $\mathcal{D}$. To reduce the number of the regions to be considered we have used the cyclic symmetry of the integral and also the identity 
\begin{equation}
P^{\left(6-2\epsilon\right)}\left(\boldsymbol{s}\right)=e^{i\pi\epsilon}\left[P^{\left(6-2\epsilon\right)}\left(-\boldsymbol{s}\right)\right]^{*}\,.
\end{equation}
following from, e.g., Feynman parametrization. Then we have only four non-equivalent regions: 
\begin{equation}
\text{I. }(-----),\quad  \text{II. }(----+),\quad  \text{III. }(---++),\quad \text{IV. }(--+-+), 
\end{equation}
where each region is marked by the list $(\sign s_1,\sign s_2,\sign s_3,\sign s_4,\sign s_5)$. 

Let us consider the analytical continuation of $\widetilde{P}^{(i)}_0$ integrals from the region $(-----)$ to the region $(----+)$. We put $s_5=|s_5|e^{i\phi}$ and change $\phi$ from $\pi$ to $0$. While changing $\phi$, we track the motion of the braniching points $t_{ai},t_{bi},t_{ci},t_{0i},t_{\infty i}$ and deform the integration contours over $t$ in such a way that they do not cross these points (and $t=0$). We should also track the changing of the argument of $F$ in the end point $t=1$. In what follows we assume, for definiteness, that $s_1<s_2<s_3<s_4<s_5$.

\begin{figure}
\begin{center}
\includegraphics[width=1\textwidth]{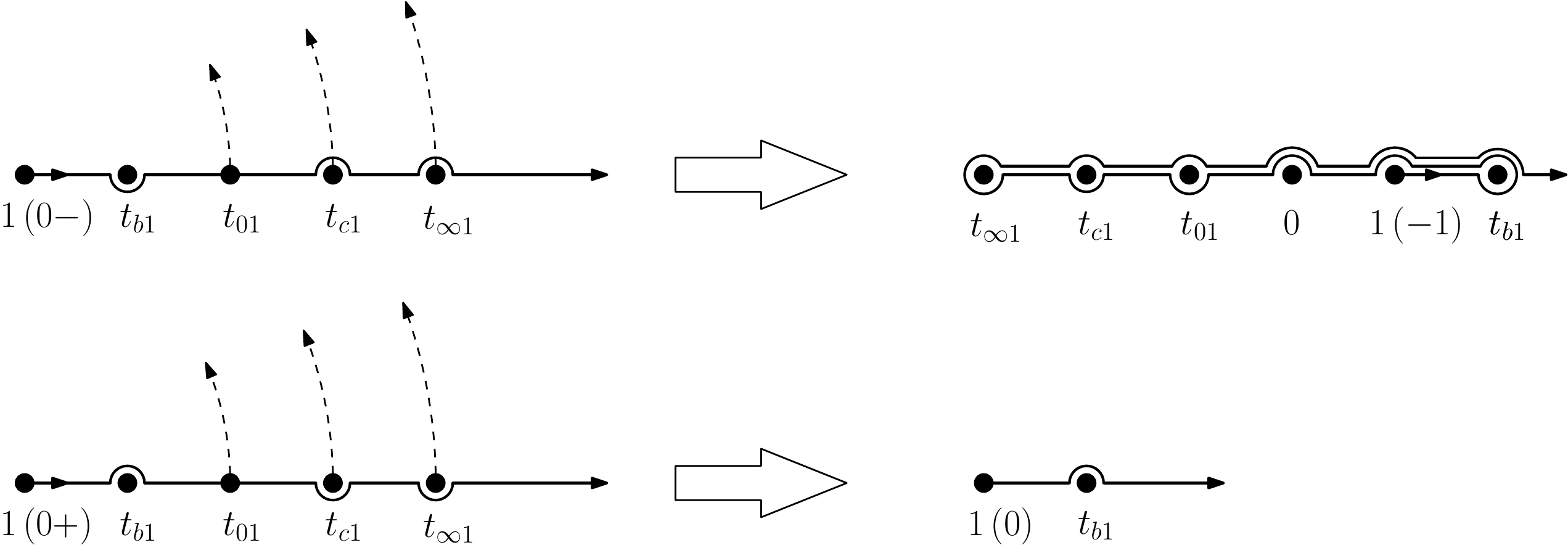}
\end{center}
\caption{Motion of the branching points of the integrand in Eq. \eqref{eq:example_integral} and the corresponding deformation of the integration contours. Upper (lower) half corresponds to the $+i0$ ($-i0$) prescription in the denominator of the argument of $f$.
Left half: $s_5<0$ ($\phi=\pi$), right half: $s_5>0$ ($\phi=0$). Dashed arrows denote the movement of the branching points upon varying $\phi$ from $\pi$ to $0$. Notation $(n\pm)$ stands for the argument lying on the $n$-th sheet on the upper/lower bank of the cut.
}\label{fig:contours}
\end{figure}

Let us explain our method on the example of the integral
\begin{equation}\label{eq:example_integral}
\frac{1}{2}\sum_{\pm}\int_1^{\infty}dt\,t^{\epsilon-1}\frac{1}{r_1(t)}F^{(0)}\biggl(\frac{[r_1(t)]^2}{D(t)\pm i0}\biggr)\,,
\end{equation}
where $r_1(t)=r_1+2s_5s_2(1-t)$ and $D(t)=\Delta\frac{S}{s_1}t-\Delta$.
In Fig. \ref{fig:contours} we show the movement of the poles of the integrand upon changing $\phi$. In the final position, when $\phi=0$, the integral is written as
\begin{multline}
\frac12\bigg\{ \intop_1^{t_{b1}}dt\,t^{\epsilon-1}\frac{F^{(-1)}}{r_1(t)}
+\intop_{t_{b1}}^{0}dt\,t^{\epsilon-1}\frac{F^{(0)}}{r_1(t)}
+\intop_{0}^{t_{c1}}dt\,(t+i0)^{\epsilon-1}\frac{F^{(0)}}{r_1(t)}
+\intop_{t_{c1}}^{t_{\infty1}}dt\,(t+i0)^{\epsilon-1}\frac{F^{(0-)}}{r_1(t)}\\
+\intop_{t_{\infty1}}^{t_{c1}}dt\,(t+i0)^{\epsilon-1}\frac{F^{(0+)}}{r_1(t)}
+\intop_{t_{c1}}^{0}dt\,(t+i0)^{\epsilon-1}\frac{F^{(0)}}{r_1(t)}
+\intop_{0}^{t_{b1}}dt\,t^{\epsilon-1}\frac{F^{(0)}}{r_1(t)}
+\intop_{t_{b1}}^{\infty}dt\,t^{\epsilon-1}\frac{F^{(0-)}}{r_1(t)}\\
+\intop_{1}^{t_{b1}}dt\,t^{\epsilon-1}\frac{F^{(0)}}{r_1(t)}
+\intop_{t_{b1}}^{\infty}dt\,t^{\epsilon-1}\frac{F^{(0-)}}{r_1(t)}\bigg\}
\end{multline}
where we suppressed the argument $[r_1(t)]^2/D(t)$ of $F^{(n)}$. The superscript $(n\pm)$ denotes the argument lying on the $n$-th sheet on the upper/lower bank of the cut. The first two lines correspond to the contribution of the upper contour and the last line corresponds to that of the lower contour in Fig. \ref{fig:contours}. Using Eq. \eqref{eq:branches}, we reduce the above expression to the form
\begin{equation}
\intop_{1}^{t_{b1}}dt\,t^{\epsilon-1}\frac{F^{(0)}}{r_1(t)}
 +\frac12\intop_{t_{b1}}^{\infty}dt\,t^{\epsilon-1}\frac{F^{(0+)}+F^{(0-)}}{r_1(t)}
 -\frac{i\pi}{2}\left[\intop_{t_{\infty1}}^{t_{c1}}+\intop_{1}^{\infty}\right]\frac{t^{\epsilon-1}dt}{\sqrt{-D(t)}}
\end{equation}
Considering in the same way all the integrals appearing in $\widetilde{P}_0^{(1-5)}$, we have
\begin{align}
\frac{\widetilde{P}_0^{(1)}(\boldsymbol{s}\in\mathcal{R})}{\sqrt{\Delta}}&=(-s_1)^{-\epsilon}\intop_1^{\infty}dt\,t^{\epsilon-1} G_1(\boldsymbol{s},t) \underline{-(-s_1)^{-\epsilon}\frac{i\pi}{2\sqrt{\Delta}}\left(\intop_{t_{\infty1}}^{\infty}\frac{dt\,t^{\epsilon-1}}{\sqrt{1-\frac{S}{s_1}t}}+\intop_{t_{\infty1}}^{t_{c1}}\frac{dt\,t^{\epsilon-1}}{\sqrt{1-\frac{S}{s_1}t}}\right)}\,,\nonumber\\
\frac{\widetilde{P}_0^{(2)}(\boldsymbol{s}\in\mathcal{R})}{\sqrt{\Delta}}&=(-s_2)^{-\epsilon}\intop_1^{\infty}dt\,t^{\epsilon-1} G_2(\boldsymbol{s},t)
\underline{-(-s_2)^{-\epsilon}\frac{i\pi}{2\sqrt{\Delta}}\left(\intop_{t_{\infty2}}^{\infty}\frac{dt\,(t+i0)^{\epsilon-1}}{\sqrt{1-\frac{S}{s_2}t}}+\intop_{t_{\infty2}}^{t_{a2}}\frac{dt\,(t+i0)^{\epsilon-1}}{\sqrt{1-\frac{S}{s_2}t}}\right)}\,,\nonumber\\
\frac{\widetilde{P}_0^{(3)}(\boldsymbol{s}\in\mathcal{R})}{\sqrt{\Delta}}&=(-s_3)^{-\epsilon}\intop_1^{\infty}dt\,t^{\epsilon-1} G_3(\boldsymbol{s},t)
\underline{-(-s_3)^{-\epsilon}\frac{i\pi}{2\sqrt{\Delta}}\left(\intop_{t_{\infty3}}^{\infty}\frac{dt\,(t+i0)^{\epsilon-1}}{\sqrt{1-\frac{S}{s_3}t}}-\intop_{t_{\infty3}}^{t_{a3}}\frac{dt\,(t+i0)^{\epsilon-1}}{\sqrt{1-\frac{S}{s_3}t}}\right)}\,,\nonumber\\
\frac{\widetilde{P}_0^{(4)}(\boldsymbol{s}\in\mathcal{R})}{\sqrt{\Delta}}&=(-s_4)^{-\epsilon}\intop_1^{\infty}dt\,t^{\epsilon-1} G_4(\boldsymbol{s},t)
\underline{-(-s_4)^{-\epsilon}\frac{i\pi}{2\sqrt{\Delta}}\left(\intop_{t_{\infty4}}^{\infty}\frac{dt\,t^{\epsilon-1}}{\sqrt{1-\frac{S}{s_4}t}}
+\intop_{t_{b4}}^{t_{\infty4}}\frac{dt\,t^{\epsilon-1}}{\sqrt{1-\frac{S}{s_4}t}}\right)}\,,\nonumber\\
\frac{\widetilde{P}_0^{(5)}(\boldsymbol{s}\in\mathcal{R})}{\sqrt{\Delta}}&=(-s_5-i0)^{-\epsilon}
\intop_1^{\infty}dt\,t^{\epsilon-1}G_5(\boldsymbol{s},t)\,,\label{eq:s5>0}
\end{align}
where
\begin{equation}
\mathcal{R}=\{\boldsymbol{s}| s_1<0,s_2<0,s_3<0,s_4<0,s_5>0\}\,.
\end{equation}
Using the relations
\begin{equation}\label{eq:s-points}
\frac{t_{ai}}{s_i}=\frac{t_{b(i+2)}}{s_{i+2}}=\frac{t_{c(i-2)}}{s_{i-2}}\,,
\end{equation}
the sum of the underlined terms in Eq. \eqref{eq:s5>0} is transformed to 
\begin{equation}
-2i\pi \frac{S^{-\epsilon}}{\sqrt{\Delta}}\int_{-1}^{\infty}\frac{(t+i0)^{\epsilon-1}dt}{\sqrt{t+1}}= -2\frac{S^{-\epsilon}}{\sqrt{\Delta}}\pi^{\frac{3}{2}}e^{i\pi\epsilon}\frac{\Gamma(1/2-\epsilon)}{\Gamma(1-\epsilon)}
\end{equation}
Note that this is exactly the second term in square brackets of Eq. \eqref{eq:pentagon_euclidean} analytically continued to the region $s_5>0$ and taken with opposite sign. Therefore, the analytical continuation of $\widetilde{P}$ to the region $\mathcal R$ has the form
\begin{multline}
\sum_{i=1}^5\frac{\widetilde{P}^{(i)}_0(\boldsymbol{s})}{\sqrt{\Delta}}+2\pi^{\frac{3}{2}}\frac{\Gamma(1/2-\epsilon)}{\Gamma(1-\epsilon)}\frac{(-S)^{-\epsilon}}{\sqrt{\Delta}}\\
=\sum_{i=1}^{4}(-s_i)^{-\epsilon}\int_1^{\infty}dt\,t^{\epsilon-1} G_i(\boldsymbol{s},t)+e^{i\pi\epsilon}s_5^{-\epsilon}\int_1^{\infty}dt\,t^{\epsilon-1}G_5(\boldsymbol{s},t)\,.
\end{multline}
Analytical continuation to other regions is performed in the same way. The outcome is that 
\begin{multline}\label{eq:tilde_pentagon}
P^{(6-2\epsilon)}=\frac{C(\epsilon)}{\epsilon}\left[\sum_{i=1}^{5}(-s_i-i0)^{-\epsilon}\int_1^{\infty}dt\,t^{\epsilon-1}G_i(\boldsymbol{s},t)
+2\pi^{\frac{3}{2}}\frac{\Gamma(1/2-\epsilon)}{\Gamma(1-\epsilon)}\frac{(-S-i0)^{-\epsilon} }{\sqrt{\Delta}}\Theta\left(s_is_j>0\right)\right]
\,,
\end{multline}
where $\Theta\left(s_is_j>0\right)$ equals to $1$ if all $s_i$ are of the same sign, and zero otherwise.
Note that the coefficient in front of $\Theta\left(s_is_j>0\right)$ has a branching point $\Delta=0$. However, when all $s_i$ are of the same sign, $\Delta$ is strictly positive. Therefore, Eq. \eqref{eq:tilde_pentagon} has no branching at $\Delta=0$\footnote{Note that $\Delta$ may vanish in regions II-IV.}.

Finally, we use relation 
\begin{equation}
\frac{r_i+2s_{i-1}s_{i+1}(1-t)}{b_{i}(t)}=
-\frac{1-\frac{r_{i+2}}{b_{i}(t)}\frac{r_{i-2}}{b_{i}(t)}}{\frac{r_{i+2}}{b_{i}(t)}+\frac{r_{i-2}}{b_{i}(t)}}
\end{equation}
and elementary trigonometric formulas to represent $G_i(\boldsymbol{s},t)$ in the form
\begin{multline}
\label{eq:G1}
G_i(\boldsymbol{s},t)=\Re\ \frac1{b_i(t)}\bigg\{
\arctan\frac{b_{i}(t)}{r_{i}}-\arctan\frac{b_{i}(t)}{r_{i+2}}-\arctan\frac{b_{i}(t)}{r_{i-2}}\\
+\frac{\pi}{2}\left[\sign r_{i+2}+\sign r_{i-2}-\sign r_{i} -\sign\left(r_{i+2}+r_{i-2}\right)\right]\bigg\}\,.
\end{multline}
Substituting Eq. \eqref{eq:G1} in Eq. \eqref{eq:tilde_pentagon}, we obtain our main result \eqref{eq:pentagon_final}.

\section{Conclusion}

In the present paper we applied the differential equation approach to the calculation of the pentagon integral $P$ in arbitrary dimension $d$. Our main result is the one-fold integral representation, Eq. \eqref{eq:pentagon_final}, valid for any real values of the invariants $s_i$. The integral in Eq. \eqref{eq:pentagon_final} converges for $d>4$ and trivially determines any order of $\epsilon$ expansion near $d=6-2\epsilon$ as a one-fold integral of elementary functions, see Eq. \eqref{eq:pentagon_final_expansion}. We have demonstrated that this integral can be expressed via the Goncharov's polylogarithms.

The simple form of the obtained result \eqref{eq:pentagon_final} hints for a possibility to find a similar representation for more complicated one-loop integrals. In particular, it would be interesting to consider the on-shell hexagon and off-shell pentagon integrals.

\acknowledgments
We are grateful to V.S. Fadin for attracting our attention to the problem of calculation of the pentagon integral and for continuing interest to our work. This work is supported in part by the RFBR grants nos. 15-02-07893 and 13-02-01023. The work of M.G. Kozlov was also supported via RFBR grant no. 16-32-60033.

\appendix
 
\section{Expansion in $\epsilon$}\label{sec:expansion}
First, we note that it is trivial to obtain any order of expansion in $\epsilon$ in terms of a one-fold integral of elementary functions from Eq. \eqref{eq:pentagon_final}. It simply amounts to writing
\begin{multline}\label{eq:pentagon_final_expansion}
\left[\frac{P^{\left(6-2\epsilon\right)}}{C(\epsilon)}\right]_{\epsilon^n}=
\Theta\left(s_is_j>0\right)\left[\frac{2\pi^{3/2}\Gamma\left[1/2-\epsilon\right]}{\Gamma\left[1-\epsilon\right]\sqrt{\Delta}}\left(-S-i0\right)^{-\epsilon}\right]_{\epsilon^{n+1}}
\\
+\sum_{i=1}^{5}\intop_{1}^{\infty}\frac{dt}{t}
\frac{\ln^{n+1}(-t/s_i+i0)}{(n+1)!}\,
\Re\ \frac1{b_i(t)}\bigg\{
\arctan\frac{b_{i}(t)}{r_{i}}-\arctan\frac{b_{i}(t)}{r_{i+2}}-\arctan\frac{b_{i}(t)}{r_{i-2}}\\
+\frac{\pi}{2}\left[\sign r_{i+2}+\sign r_{i-2}-\sign r_{i} -\sign\left(r_{i+2}+r_{i-2}\right)\right]
\bigg\}\,,
\end{multline}
where $\left[f(\epsilon)\right]_{\epsilon^n}$ denotes the coefficient in front of $\epsilon^n$ in the expansion of $f(\epsilon)$ in $\epsilon$.

Let us explain how to obtain the expansion of $P^{(6-2\epsilon)}$ in terms of generalized polylogarithms. We restrict ourselves by the Euclidean region. In order to express the results in a compact form, we introduce the notation $a{\pm}$ for the integration weights
\begin{equation}
w(a+,x)=\frac{2a}{x^2-a^2}\,\quad w(a-,x)=\frac{2x}{x^2-a^2}
\end{equation}
These weights are simply the linear combinations of the conventional weights $w(a,x)=\frac{1}{x-a}$:
\begin{equation}
w(a\pm,x)=w(a,x)\mp w(-a,x)\,.
\end{equation}
We define, as usual, see, e.g., Ref. \cite{Duhr2012}, the iterated integrals
\begin{equation}
G(a_1,a_2,\ldots|y)=\intop_0^y dx\, w(a_1,x)\, G(a_2,\ldots|x)\,.
\end{equation}
In Euclidean region $\Delta$ is always positive, and it is convenient to use the variables $a_{i}=r_{i}/\sqrt{\Delta}$, which satisfy
\begin{gather}
\sum_{i} a_ia_{i+2}=1\,, \\
a_i>-1\,,\quad a_i+a_{i+1}>0\,.
\end{gather}
Pulling out the overall factor $\frac{\left(-S\right)^{-\epsilon}}{\sqrt{\Delta}}$, we obtain
\begin{multline}\label{eq:euclideanRegion}
\frac{P^{\left(6-2\epsilon\right)}\left(s_{1},\,s_{2},\,s_{3},\,s_{4},\,s_{5}\right)}{\frac{2\Gamma(1-\epsilon)^{2}\Gamma(1+\epsilon)}{\Gamma(1-2\epsilon)\epsilon\sqrt{\Delta}\left(-S\right)^{\epsilon}}}=\sum_{i=1}^{5}\left[
T(a_i,y_i)-T(a_{i+2},y_i)-T(a_{i-2},y_i)\right]
+\frac{2\pi^{3/2}\Gamma\left[1/2-\epsilon\right]}{\Gamma\left[1-\epsilon\right]}\,,
\end{multline}
where $y_{i}=\sqrt{S/s_{i}-1}$ and the function $T$ are defined as
\begin{equation}
T\left(a,y\right)=\Re\intop_{1+y^{2}}^{\infty}dt\ t^{\epsilon-1}\frac{1}{\sqrt{t-1}}\left[\frac{\pi}{2}-\arctan\frac{a}{\sqrt{t-1}}\right]\,.
\end{equation}
Note that replacing in this formula $\frac{\pi}{2}-\arctan \frac{a}{\sqrt{t-1}}$ with $\arctan\frac{\sqrt{t-1}}{a}$ is not valid for $a<0$.
When the second argument of the function $T$ is zero, the integral can be taken in terms of generalized hypergeometric functions
\begin{equation}\label{eq:T(0)}
T\left(a,0\right)=\frac{\pi ^{3/2} \theta (-a) \Gamma \left(\frac{1}{2}-\epsilon \right)}{\Gamma (1-\epsilon )}-\frac{\, _3F_2\left(\frac{1}{2},1,1;\frac{3}{2},1+\epsilon ;\frac{1}{a^2}\right)}{a\epsilon}-\frac{\pi  |a| ^{2 \epsilon } \, _2F_1\left(\frac{1}{2}-\epsilon ,1-\epsilon ;\frac{3}{2}-\epsilon ;\frac{1}{a^2}\right)}{a(2 \epsilon -1) \sin(\pi  \epsilon ) }
\end{equation}
These functions can be readily expanded using standard tools, like \texttt{HypExp}, \cite{Huber2006}. 
In order to expand the difference $T\left(a,y\right)-T\left(a,0\right)$, we pass to the variable
$\tau=\sqrt{t-1}$ and expand under the integral sign:
\begin{equation}
T\left(a,y\right)-T\left(a,0\right)=-\sum_{n=0}^{\infty}\epsilon^n\Re\intop_0^{y}d\tau\ \frac{2}{1+\tau^2} \frac1{n!} \ln^n (1+\tau^2)\left[\frac{\pi}{2}-\arctan\frac{a}{\tau}\right]\,.
\end{equation}
Taking into account that 
\begin{gather*}
\frac{\ln^n(1+\tau^2)}{n!} =G(\{i-\}_n|\tau)\stackrel{\text{def}}{=}G(\underbrace{i-,\ldots,i-}_n|\tau)\,,\\
\pi/2 -\arctan\frac{a}{\tau} =\pi\theta(-a) -i G(ia{+}|\tau)\,,
\end{gather*}
and using shuffling relations, we obtain  
\begin{equation}\label{eq:T(y)-T(0)}
T\left(a,y\right)-T\left(a,0\right)=\sum \epsilon^n\Re\left\{[G(ia{+} |y)+i\pi\theta(-a)]G(i+,\{i-\}_n|y)-G(ia{+},i+,\{i-\}_n|y)\right\}\,.
\end{equation}
Equations \eqref{eq:T(0)} and \eqref{eq:T(y)-T(0)} allow one to obtain any term of expansion of the pentagon integral near $d=6$. In order to obtain the expansion of the integral near $d=4$, one may use the dimensional recurrence relation \eqref{eq:dre}.

The pentagon integral is finite in $d=6$, therefore, the $1/\epsilon$ term should vanish. The cancellation of the divergencies in individual terms in Eq. \eqref{eq:euclideanRegion} is quite tricky. First, we note that 
\begin{multline}
\left.T\left(a,y\right)\right|_{\epsilon=0}=T_0\left(a,y\right)=\Re\bigg[\frac{1}{2} \text{Li}_2\left(\frac{(a-1) (y+i)}{(a+1) (y-i)}\right)+\frac{1}{2} \text{Li}_2\left(\frac{(a-1) (y-i)}{(a+1) (y+i)}\right)-\text{Li}_2\left(\frac{a-1}{a+1}\right)\\
-\arctan^2 y+\frac{\pi ^2}{4}\bigg]
\end{multline}
We want to prove that 
\begin{equation}\label{eq:cancellation}
\sum_{i=1}^{5}\left[
T_0(a_i,y_i)-T_0(a_{i+2},y_i)-T_0(a_{i-2},y_i)\right]
+2\pi^{2}=0
\end{equation}
in the whole Euclidean region. Let us first show that the left-hand side is constant. The differential of the left-hand side is
\begin{multline}
\sum_{i=1}^{5}\Re\bigg\{
\frac{2dy_i}{1+y_i^2}\left[\frac{\pi}2+\arctan\frac{a_i}{y_i}-\arctan\frac{a_{i+2}}{y_i}-\arctan\frac{a_{i-2}}{y_i}\right]\\
+\log  \left(\frac{y_i^2+a_i^2}{y_i^2+1}\right) \frac{da_i}{1-a_i^2}
-\log \left(\frac{ y_i^2+a_{i+2}^2}{y_i^2+1}\right) \frac{da_{i+2}}{1-a_{i+2}^2}
-\log \left(\frac{y_i^2+a_{i-2}^2}{y_i^2+1}\right) \frac{da_{i-2}}{1-a_{i-2}^2}
\bigg\}
\end{multline}
The differential $dy_i$ can be expressed via $da_i$,  $da_{i+2}$,  $da_{i-2}$, but we may refrain from doing it thanks to the following remarkable fact:
the quantity $\Re y\left[\frac{\pi}2+\arctan\frac{a}{y}-\arctan\frac{b}{y}-\arctan\frac{c}{y}\right]$ vanishes after the substitution $y=\sqrt{bc-ab-ac}$. Then, the coefficient in front of $\frac{da_i}{1-a_i^2}$ becomes 
\begin{equation}
\Re \log \frac{ \left(a_i^2+y_i^2\right)\left(y_{i+2}^2+1\right)\left(y_{i-2}^2+1\right)}{\left(y_i^2+1\right)\left(a_i^2+y_{i+2}^2\right)\left(a_i^2+y_{i-2}^2\right)} 
\end{equation}
Substituting $y_{i}=\sqrt{S/s_{i}-1}$ and $a_i=r_i/\sqrt{\Delta}$, we verify that this coefficient is zero.
Therefore, in order to prove the identity \eqref{eq:cancellation}, we need to calculate the left-hand side in any specific point $(s_1,s_2,s_3,s_4,s_5)$ in the Euclidean region. We choose symmetric point $s_1=s_2=\ldots=s_5=-1$, where $a_k=\frac{1}{\sqrt{5}}$ and $y_k=\frac{i}{\sqrt{5}}$. Then
\begin{multline}
T_0\left(\frac{1}{\sqrt{5}},\frac{i}{\sqrt{5}}\right)=
\frac{1}{2} \text{Li}_2\left(\left(\frac{\sqrt{5}-3}{2}\right)^2\right)-\text{Li}_2\left(\frac{\sqrt{5}-3}{2}\right)+\frac{\pi ^2}{3}+\arctanh^2\left(\frac{1}{\sqrt{5}}\right)\\
=\text{Li}_2\left(\frac{3-\sqrt{5}}{2}\right)+\frac{\pi ^2}{3}+\arctanh^2\left(\frac{1}{\sqrt{5}}\right)=\frac{2 \pi ^2}{5}\,.
\end{multline}
The last transition is due to one of the eight remarkable values of dilogarithm, see, e.g., Ref. \cite{zagier2007dilogarithm}. Using this identity, it is easy to see that Eq. \eqref{eq:cancellation} holds in the symmetric point, and, therefore, in the whole Euclidean region. Similar analysis shows the cancellation of $\epsilon^{-1}$ terms in Eq. \eqref{eq:pentagon_final} in all regions.

\bibliographystyle{JHEP}
\bibliography{pentagon}

\end{document}